\documentclass[twocolumn,twoside,slac_two]{revtex4}
\usepackage{graphicx}
\usepackage{fancyhdr}
\begin{document}

%Title of paper
\title{Gadolinium study for a water Cherenkov detector}

% Repeat the \author .. \affiliation  etc. as needed
%
% \affiliation command applies to all authors since the last
% \affiliation command. The \affiliation command should follow the
% other information

\author{Atsuko Kibayashi for the Super-Kamiokande Collaboration}
\affiliation{Department of Physics, Okayama University, Okayama, Japan}

\begin{abstract}
Modification of large water Cherenkov detectors by addition of gadolinium has 
been proposed. The large cross section for neutron capture on Gd will greatly 
improve the sensitivity to antielectron neutrinos from supernovae and reactors.
A five-year project to build and develop a prototype detector based on 
Super-Kamiokande (SK) has started. We are performing various studies, 
including a material soak test in Gd solution, light attenuation length 
measurements, purification system development, and neutron tagging efficiency 
measurements using SK data and a Geant4-based simulation. 
We present an overview of the project and the recent R\&D results.
\end{abstract}

%\maketitle must follow title, authors, abstract
\maketitle

\thispagestyle{fancy}

% body of paper here - Use proper section commands
% References should be done using the \cite, \ref, and \label commands
% Put \label in argument of \section for cross-referencing
%\section{\label{}}

%%%%%%%%%%%%%%%%%%%%%%%%%%%%%%%%%%
\section{Introduction}
Super-Kamiokande (SK) \cite{SK} is the largest light water
Cherenkov detector that has been successfully
observing solar, atmospheric and accelerator neutrinos.
Recently the addition of 0.2\% gadolinium in the SK detector 
has been proposed~\cite{Gd-SK}.
This modification can greatly improve the detection
sensitivity of anti-electron neutrinos.
The interaction in the water $\bar{\nu}_e + p \to e^+ + n$
emits a positron and a neutron.
The positron radiating Cherenkov photons
is immediately detected.
The neutron is quickly thermalized in the water,
and is captured by Gd with a probability of 90\% within 20~$\mu$s and 4~cm.
The neutron captured Gd emits 3-4 gamma rays
having the total energy of about 8~MeV.
The time and spatial correlation of the positron and neutron captured
events can significantly reduce the backgrounds, and hence
enhance the $\bar{\nu}_e$ signal events. 

The modification of SK enables us to increase
the detectability of the supernova relic neutrinos (SRN).
It is estimated that about $10^{17}$ supernovae have happened
since the first star formation in the universe.
All the past supernovae ejected not only the heavy elements
but also huge numbers of neutrinos.
Therefore, the measurements of the SRN flux and energy spectrum
provide the information of history of the heavy element production,
combined with the supernova neutrino generation mechanism.
So far, the SK experiment has given the most stringent limit on the
SRN flux: 1.2 $\bar{\nu}_e$~cm$^{-2}$s$^{-1}$ at the 90\%
confidence level with the energy threshold of 19.3~MeV~\cite{SK-SRN}.
The flux limit is three times larger than the theoretical predictions.
Thus we need the sensitivity improvement with a factor of three. 

The energy threshold was set to optimize the
signal to background ratio.
The dominant background is the radioactive spallation nuclei
that are constantly created in the water
by collisions on the oxygen nuclei with high energy
comic ray muons.
The background level steeply increases as the
threshold energy decreases.
Another serious background is the invisible muons
produced via the interactions of atmospheric neutrinos
$\nu_{\mu} + \mbox{N} \to \mu + \mbox{N}^{\prime}$.
The muon is invisible because its energy is lower than
the Cherenkov threshold, and hence only the decay electron is
visible.
Although the event rate can be estimated from data
using decay electron energy spectrum shape,
the statistical uncertainty of the number of events
decreases the SRN sensitivity.
Because of the presence of the backgrounds,
the current experimental sensitivity will not be improved;
we need 40 years' operation to lower the flux limit by a factor of three. 

Making use of the neutron tagging with Gd
can significantly reduce both spallation events and invisible muons;
the energy threshold can be lowered to 10~MeV
and the invisible muon events can be reduced by
a factor of five.
The model dependent expected SRN event rate
is 0.8 to 5.0 events per year
in the 22.5~kton fiducial volume in the energy
range of 10-30~MeV~\cite{SRN-model}.
Five years' operation of SK with Gd
can easily achieve the required sensitivity improvement. 

By lowering the energy threshold further down to 2.5~MeV,
SK with Gd could detect the reactor neutrinos
with the best precision, allowing the accurate measurements
of the neutrino oscillation parameters.
The neutron tagging can be used to distinguish the interactions
between neutrinos and anti-neutrinos that preferably emit protons
and neutrons, respectively. This technique is important
for neutrino oscillation studies with
atmospheric and accelerator neutrinos. 

As mentioned above,
the Gd loaded SK has a possibility to open the new era
of the neutrino physics.
On the other hand, a lot of research and development (R\&D) studies
have to be conducted:
studies on the chemical reactions of the Gd solution
to the material components comprised in SK,
the Gd solution transparency measurements,
the water purification system,
measurements of the ambient neutron background fluxes
and measurements of the neutron tagging efficiency.
We have already started those studies.
In this paper we present the status of the R\&D.
Finally we will mention the current plan
to construct a test facility consisting of
a 200 ton Gd solution Cherenkov detector,
a transparency measurement instrument
and a water purification system.
The test tank detector performance is studied
using a Geant4 based Monte-Carlo (MC) simulation.

%The first step to preparing the paper is to copy the slac\_two.rtx file, 
%available at the 
%DPF-2009 web site\footnote{http://www.dpf2009.wayne.edu/proceedings.php} 
%for download as a template package, into a new directory.
%Next copy this template (dpf2009\_template.tex) 
%and rename it as dpf2009\_SpeakerName.tex, 
%where SpeakerName should be replaced by the author name
%(remove spacing between the first and last names as in this example). 
%Edit the file to make the necessary modifications and save when finished. 
%
%Because this template has been set up to meet requirements for conference proceedings  
%papers, it is important to maintain these established styles.   
%Other editorial guidelines are described in the next section.

%%%%%%%%%%%%%%%%%%%%%%%%%%%%%%%%%%
\section{The R\&D status and results}

%%%%%%%%%%%%%%%%%%%%%%%%%%%%%%%%%%
\subsection{The soak tests with Gd solutions}
The candidates of the Gd compounds that can be
resolved in water are
GdCl$_3$, Gd$_2$(SO$_4$)$_3$ and Gd(NO$_3$)$_3$.
It has been shown that the transparency of the GdCl$_3$ doped 
water in stainless steel lined tank decreases rapidly in the UV wavelength 
region~\cite{svoboda}.
The soak test studies of stainless-steel samples
in the GdCl$_3$ solution were also  performed
with the accelerating condition keeping the solution
temperature at 60~$^{\circ}$C.
Cracks and corrosion have been found
for some samples which are stressed and thermally activated
artificially.
For those samples, we simulate the welding of the stainless
steel support structure in the SK tank.
On the other hand, no damage is found
for the solution of
Gd(NO$_3$)$_3$ and Gd$_2$(SO$_4$)$_3$~\cite{mitsui}.
However in the Gd(NO$_3$)$_3$ solution, strong absorption 
by Nitrate was evident below 350 nm, in the 
wavelength region matched with those of Cherenkov photons.
Thus, gadolinium sulfate Gd$_2$(SO$_4$)$_3$
is found to be the best candidate so far.
We have started the soak test with the gadolinium sulfate
solution for 37
material components currently used in SK, including
the stainless steel, PMT support rubber,
plastic tyvek, acrylic material, cable and so on.
For large size components,
we have cut those to small samples of 3$\times$3~cm$^2$.
Each sample is encapsulated in a 500~ml plastic bottle
with the Gd solution or pure water after
nitrogen gas bubbling to get rid of oxygen.
The bottles are left in room temperature for three months.
No damage has been found for all the samples so far.
Now we are measuring the concentration of any
chemical contamination dissolved into the Gd solution.

%%%%%%%%%%%%%%%%%%%%%%%%%%%%%%%%%%
\subsection{The water transparency measurements}
\begin{figure}
\centering
\includegraphics[width=3.5cm, angle=180]{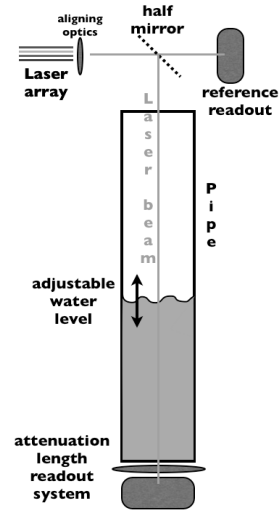}
\caption{The water transparency measurement facility at UCI.}
\label{fig:transparency}
\end{figure}

The light attenuation length in the water is the
key quantity that determines the performance
of the water Cherenkov detector.
The current SK experiment measures the attenuation length
using natural sources of cosmic ray muons
and muon decay electrons, and artificial
light sources such as the laser.
In order to continue the various physics programs
in SK with the Gd-loaded water,
the attenuation length must be kept larger than 70~m,
the level in the current SK.
The water transparency measurement facility
has been constructed at University of California, Irvine (UCI).
Figure~\ref{fig:transparency} depicts the facility.
An array of lasers that emits a light beam with
six different wavelengths from 337~nm to 650~nm is
placed on the top.
The beam is reflected by a half mirror and injected
to a pipe containing water or Gd solution to measure.
The half of the beam goes through the half mirror and
is detected by PIN photo-diodes to monitor the light
intensity pulse by pulse.
The pipe diameter and length are 16.5~cm and 6.3m, respectively.
The intensity of the light beam injected to the pipe
is also monitored by PIN photo-diodes
at the bottom of the pipe.
By changing the water level in the pipe,
we measure the light attenuation.
The first measurement using pure water was performed.
The measured length is within the expectations.
Currently, measurements in the Gd$_2$(SO$_4$)$_3$ doped water are
being carried out.

\begin{figure}
\centering
\includegraphics[width=5.5cm, angle=-90]{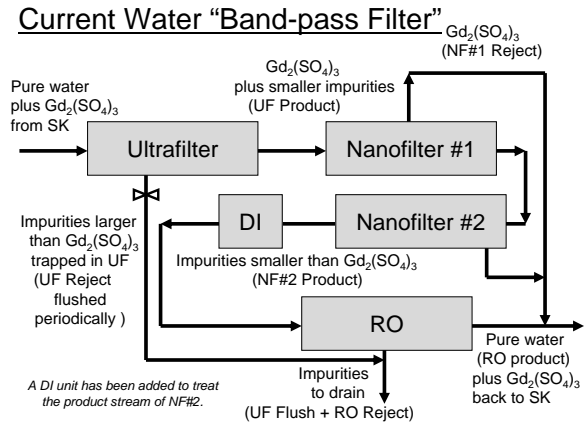}
\caption{The proposed water purification system. }
\label{fig:water-purification}
\end{figure}

%%%%%%%%%%%%%%%%%%%%%%%%%%%%%%%%%%
\subsection{The water purification system}
The water purification system employed in the present SK
rejects all the contamination in the water.
Without modification, Gd and SO$_4$ would also be
removed from the detector.
A selective filtration system has been proposed
to prevent that~\cite{purification}.
Figure~\ref{fig:water-purification} shows
the proposed system.
The SK water containing gadolinium sulfate is fed to
the ultrafilter, where all the contamination
of which size is larger than the size of Gd and SO$_4$
is removed.
The water with Gd$_2$(SO$_4$)$_3$ plus smaller impurities
is sent to the two nanofilters.
Since the nanofilter hole size is smaller than Gd and SO$_4$,
almost all the gadolinium sulfate is removed.
The water containing impurities smaller than Gd and SO$_4$ is 
deionized (DI) and then sent through  
the reverse osmosis (RO), where only H$_2$O molecules 
can pass through the membrane.
The pure water from the RO and the gadolinium sulfate solution
from the nanofilters are combined to be sent to the 
Cherenkov detector. 

The prototype of the selective filtration system was
constructed at UCI.
%Figure~\ref{fig:water-system-uci} shows a picture of the system.
%The Gd and SO$_4$ rejection efficiency at the nanofilters
%are measured to be greater than 99.99\%.
%While the output pure water from the RO contain Gd and SO$_4$
%less than measurable limits of 0.006\% and 0.11\%, respectively.
%Therefore the prototype system has demonstrated that
%the selective filtration can work very well.

%\vspace{0.5cm}
%\begin{figure}[h]
%\centering
%\includegraphics[width=6.cm, angle=-90]{dpf2009_fig03.eps}
%\caption{The water purification system at UCI. }
%\label{fig:water-system-uci}
%\end{figure}

%%%%%%%%%%%%%%%%%%%%%%%%%%%%%%%%%%
\subsection{Neutron tagging efficiency measurements at SK}
\begin{figure}[h]
\centering
\includegraphics[width=7.cm]{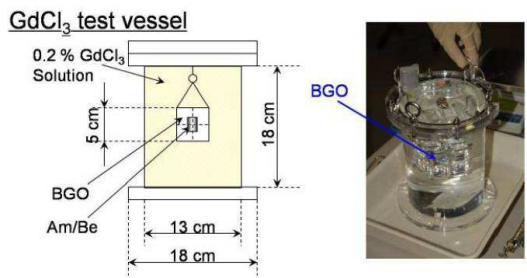}
\caption{The test vessel used for the neutron tagging efficiency
measurements in SK. }
\label{fig:vessel}
\end{figure}

We have performed the neutron tagging efficiency measurements
at SK using a test vessel
as shown in Fig.~\ref{fig:vessel}~\cite{n-tag}.
The diameter and height of the vessel is 18~cm.
The vessel frame is made of acryl.
In the vessel a BGO scintillator is placed at the center.
In the BGO scintillator, an Am/Be source is incorporated,
where the $^{241}$Am alpha decays and a reaction
$\alpha + ^9\mbox{Be} \to ^{12}\mbox{C}^{*}+n$ happens.
The excited carbon $^{12}$C$^{*}$ emits a 4.43~MeV gamma ray.
Therefore a gamma ray and a neutron are simultaneously emitted.
The gamma ray is detected by the BGO scintillator,
generating the prompt trigger in SK.
This prompt event simulates the positron from the
inverse beta reaction.
The space between the BGO scintillator and the acrylic wall of the
vessel is filled with 2.4 liters of 0.2\% GdCl$_3$ solution.
The liberated neutron is quickly thermalized in the Gd solution
and then captured by Gd.
Three to four gamma rays per one neutron capture are emitted
to the SK tank, and detected as the Cherenkov light
radiated from the Compton-scattered electrons.
The event data were collected using the delayed coincidence
trigger condition.
The vessel was placed at various positions in the SK detector.

\vspace{0.5cm}

\begin{figure}[h]
\centering
\includegraphics[width=6.5cm]{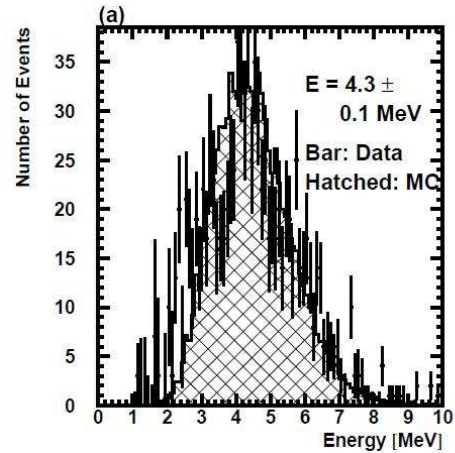}
\caption{The energy spectrum of the gamma rays from Gd
capturing a neutron. The points with bars represent data,
while the histogram shows the MC prediction.}
\label{fig:gd-gamma-energy}
\end{figure}

Figure~\ref{fig:gd-gamma-energy} shows the
measured energy distribution of the gamma rays from the Gd,
in good agreement with the MC prediction indicated as
the hatched histogram.
The mean energy is about 4.5~MeV, which is lower than
the total emission energy of 8~MeV.
This is because only the part of the energy is translated
to the Cherenkov photons due to the Compton scattering.
The neutron tagging efficiency is estimated to be 66.7\%
with a 3~MeV energy threshold, taking into account
the standard SK event reduction (80\%)
and the Gd capture efficiency (90\%) in the solution.
The accidental background rate is estimated to be $2\times 10^{-4}$
with the energy threshold of 10~MeV for the prompt $\bar{\nu}_e$ events.
The time interval between the prompt and neutron capture
events are measured.
An exponential function is fitted to the distribution,
and we obtain the lifetime of the thermal neutron of
$20.7\pm5.5$~$\mu$s, in good agreement with the
MC prediction of $20.3\pm4.1$~$\mu$s.

This study has clearly demonstrated that
the SK detector with Gd can detect the neutron with the high efficiency
and the low background level, sufficient to achieve the
required sensitivity improvement.

\vspace{0.5cm}

\vspace{0.5cm}
\begin{figure}[h]
\centering
\includegraphics[width=5cm]{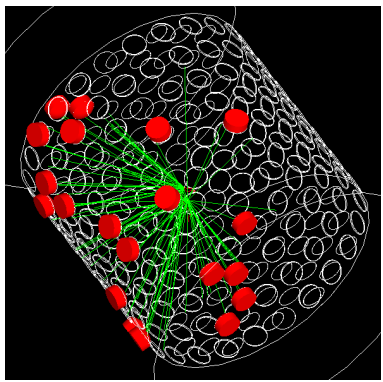}
\caption{
A typical Monte-Carlo neutron capture event.  
}
\label{fig:n_event}
\end{figure}

%%%%%%%%%%%%%%%%%%%%%%%%%%%%%%%%%%
\section{R\&D test facility construction}
\begin{figure}[h]
\centering
\includegraphics[scale=0.12]{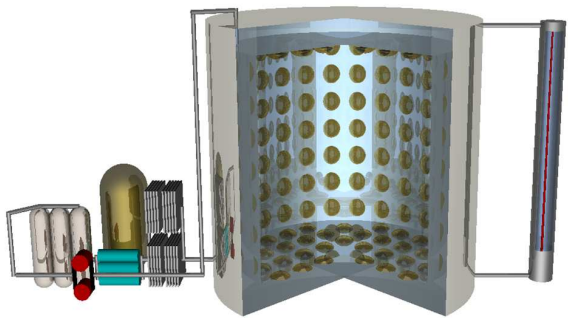}
\caption{A planned test facility for the Gd Cherenkov detector
demonstration. 
\textit{Left to right:} 
A water purification system; a cherenkov detector with 
photo-multiplier tubes; a water transparency measurement system. 
}
\label{fig:test-tank}
\end{figure}

At present, the R\&D studies mentioned above are performed individually.
To verify the neutron detection principle and
to measure the ambient neutron background rate,
we have started the construction of a test facility 
containing the water systems and a small scale prototype 
water Cherenkov detector. 
Full budget for this facility is approved. 
Laboratory space of about 15~m $\times$ 10~m $\times$ 9~m 
will be excavated in the Kamioka mine. 
The size is large enough to contain a cylindrical tank of 
size 6.5~m (diameter) $\times$  6.2~m (height), 
a 6.3~m water tower for the transparency measurement with 
a space for a worker on the top, and 
the water purification system.   
The water Cherenkov detector will 
be constructed with the same materials used in the SK detector 
including 240 20-inch PMTs.  
Figure~\ref{fig:test-tank} shows an overview of the facility.
The reconstructed vertex resolution of the Gd neutron capture
events is estimated to be about 1.5 m with a Geant4 
based simulation. This resolution is 
sufficient to the measurements for 
the ambient neutron
background rate and the neutron tagging efficiency.
Studies with the test facility is planned to continue for five years.
Gd will be loaded to SK after all the R\&D studies are
concluded.

%%%%%%%%%%%%%%%%%%%%%%%%%%%%%%%%%%
\section{Summary}
The SK detector with Gd solution has been studied.
The ability of tagging neutron can greatly improve the
sensitivity of the anti-electron neutrino detection,
enabling us to reach the theoretical predictions of the
SRN flux.
Several R\&D studies are being performed.
The best compound known so far is gadolinium sulfate
Gd$_2$(SO$_4$)$_3$.
No material damage with the solution is found.
The attenuation length measurements of the Gd solution are
in progress.
The prototype of the selective filtration system demonstrates
the Gd solution can be purified without loss of Gd.
The studies with the test vessel in SK have verified
the neutron capture event can be detected by SK
with the high neutron tagging efficiency and the low background level.
The test facility composed of the purification, transparency
measurement systems and a 200 ton water Cherenkov detector will be 
constructed in the Kamioka mine. 
Once the prototype detector has proven to work as expected after these 
R\&D studies, modification of SK with the Gd option will be proposed.

\bigskip % extra skip inserted
% Create the reference section using BibTeX:
%\bibliography{basename of .bib file}
%\begin{thebibliography}{9}   % Use for  1-9  references

\end{document}